\documentclass[%
 reprint,
 amsmath,amssymb,
 aps,
 pra,
 longbibliography,
]{revtex4-2}

\usepackage{graphicx}
\usepackage{dcolumn}
\usepackage{bm}

\usepackage{braket}
\usepackage{mathtools}
\usepackage{float}
\begin{document}

\preprint{APS/123-QED}

\title{High-Fidelity Inter-Species Rydberg Gates with Two-Photon Driving}

\author{Shuan Wang}
\author{Kevin Singh}%
 \email{kevinsingh@physics.osu.edu}
\affiliation{%
 The Ohio State University
}

\date{\today}

\begin{abstract}

Recently, neutral atom arrays have become known for their promising potential in quantum information processing. High-fidelity entangling operations are typically achieved by exciting atoms to high-lying Rydberg states through two-photon processes. To improve the performance of these systems at the hardware level, dual-species strategies have been applied to enhance readout, error correction, and reloading in neutral atom systems. However, the influence of laser crosstalk arising from two-photon excitation for dual-species systems has remained largely unexplored in the literature. Here, we develop a model that incorporates the influence of crosstalk in inter-species Rydberg gates with two-photon driving. We demonstrate that the influence of this crosstalk can be suppressed to arbitrarily small levels through appropriate gate design, with gate fidelities greater than $99.8\%$ achievable with conservative experimental parameters in the near-detuned case of an $^{85}$Rb/$^{87}$Rb system. We also outline how this fidelity can be further enhanced through improvements to the Rydberg excitation.


\end{abstract}

\maketitle


\section{Introduction}

Neutral atom arrays have recently emerged as a leading platform for quantum computation and simulation. These devices incorporate all-to-all qubit connectivity in reconfigurable architectures \cite{2ym8-vs82, Bluvstein2022, bluvstein2024logical} with coherent control over hundreds to thousands of qubits \cite{Scholl2021, Pause:24, PhysRevA.105.063111, Ebadi2021}, allowing increasingly complex Hamiltonians and circuits to be implemented and explored \cite{chen2023continuous, shaw2024multi}. 

The neutral atom array platform is developing rapidly, with researchers pursuing differing and equally compelling approaches to address current challenges on the road to fault-tolerant operation. Current challenges include: fast reloading during system operation to realize deep circuits \cite{Li2026-ql,Chiu2025-eg}, low-loss and high-speed rearrangement to realize efficient quantum-error correcting codes \cite{bhardwaj2026highrateqldpcprocessors}, mid-circuit readout for stabilizer measurements \cite{singh2023mid,PhysRevLett.129.203602,graham2023midcircuit,lis2023midcircuit,norcia2023midcircuit,bluvstein2024logical}, and efficient decoding and feedforward. A critical requirement shared by these challenges is that they need to be performed without decoherence of data qubits and with low crosstalk, especially during operations on auxiliary qubits used in ancilla-based quantum error correction. One promising strategy to achieve this requirement is dual-species atom arrays \cite{singh2023mid, Anand2024, beterov2015rydberg, zeng2017entangling, sheng2022defect, PhysRevX.12.011040, miles2026qubitsyndromemeasurementshigh}, where the unique energy level structures of each species can allow independent control as well as measurement. Recent experimental work has explored the simultaneous use of two different elements \cite{singh2023mid, beterov2015rydberg, PhysRevX.12.011040, Anand2024, miles2026qubitsyndromemeasurementshigh} or two different isotopes of the same element \cite{zeng2017entangling, sheng2022defect, PhysRevX.14.041062}. Compared with a dual-element strategy, a dual-isotope strategy allows shared optical systems \cite{Anand2024, zeng2017entangling}, significantly reduces laser-system complexity, and provides access to inter-isotope resonant Rydberg interactions. However, as of today, the fidelity of experimentally demonstrated inter-species Rydberg gates remains limited. For example, the dual-element gate in \cite{miles2026qubitsyndromemeasurementshigh} reaches a fidelity of 0.975, while the dual-isotope gate in \cite{zeng2017entangling} reaches a fidelity of 0.73(1). Both are above standard surface-code error thresholds \cite{PhysRevA.86.032324}.

\begin{figure}[b]
\includegraphics[width=\linewidth]{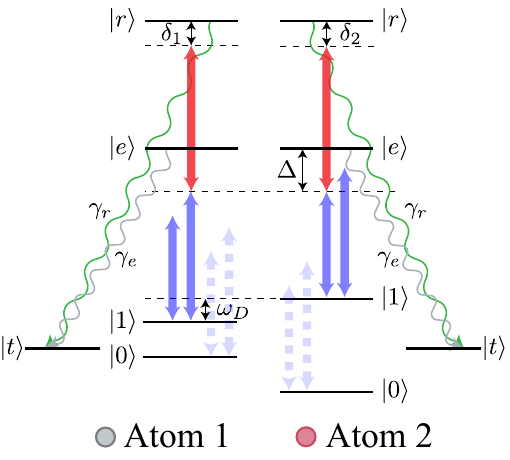}
\caption{Energy level diagrams for atom 1 and atom 2. For each atom, five energy levels $\ket{0}$, $\ket{1}$, $\ket{e}$, $\ket{r}$, $\ket{t}$ are considered. Two optical fields are used for excitation of the atoms from $\ket{1}$ to $\ket{r}$: a global red laser (solid red arrows) and a global blue laser with two frequency components (solid blue arrows). The two frequency components in the blue laser also act off-resonantly on qubit state $\ket{0}$ (dashed light blue arrows).}
\label{fig:levels}
\end{figure}

One of the main challenges for neutral-atom array systems has been improving the fidelity of entangling gates, such as the two-qubit controlled-Z (CZ) gate, based on Rydberg interactions. A compelling strategy for implementing high-fidelity gates is to design them using quantum optimal control \cite{Evered2023}. As a toolbox for crafting the shapes of external fields to accomplish given tasks with high fidelity, quantum optimal control has been widely used in many different experimental platforms related to quantum technology, including superconducting circuits \cite{Werninghaus2021}, nitrogen-vacancy centers \cite{rembold2020introduction}, and nuclear magnetic resonance \cite{PhysRevLett.118.150503}.

So far, the discussion of high-fidelity gates based on quantum optimal control in neutral-atom platforms has been primarily limited to single-species arrays \cite{Evered2023, Jandura2022}. This article illustrates the design of a two-qubit controlled-Z gate within a dual-species array where excitation to the Rydberg states occurs through a two-photon process for each atomic qubit. We evaluate the performance of such a gate based on an $^{85}$Rb/$^{87}$Rb array. The dual-isotope gate presented here only requires two lasers to achieve a high-fidelity gate between rubidium isotopes, one infrared laser and a frequency-modulated blue laser, greatly simplifying the complexity associated with the synchronization of laser pulses. Moreover, these results show that it is possible to achieve high-fidelity entangling gates between different species of atoms even when the energy levels of the species are nearly resonant (as is the case for $^{85}$Rb and $^{87}$Rb). Importantly, even in this nearly resonant case for differing isotopes, the gate does not require individual addressing of the isotopes to suppress crosstalk and can achieve high-fidelity operation beyond that of traditional so-called $\pi$---2$\pi$---$\pi$ gates \cite{PhysRevLett.85.2208} demonstrated for dual-isotope arrays \cite{zeng2017entangling}. 

\section{Theoretical Model}

\subsection{Level Structure Setup}
We consider a two-atom system, where each atom is modeled as a five-level system as shown in  Fig.~\ref{fig:levels}. The levels here are 

\begin{itemize}
\item $\ket{0}$ Qubit level 0
\item $\ket{1}$ Qubit level 1
\item $\ket{e}$ Intermediate state
\item $\ket{r}$ Rydberg state
\item $\ket{t}$ Trash state,
\end{itemize}
where the Trash state models population that has decayed out of the qubit subspace from the intermediate or excited Rydberg states.

\subsection{Hamiltonian and Pulses}

A two-photon scheme is used to excite the atoms to Rydberg states. A red laser corresponding to the transition between $\ket{e}$ and $\ket{r}$ is shared for the two atoms, while a blue laser with two frequency components is used to address the transition between $\ket{1}$ and $\ket{e}$ in both atoms, each frequency component addressing a specific hyperfine level in one isotope. The lasers are assumed to be global and overlap with both atoms. This necessitates considering the effect of both blue frequency tones on each atom as illustrated in Fig.~\ref{fig:levels}. The two frequency components in the blue laser also act off-resonantly on qubit state $\ket{0}$, potentially causing undesired far-detuned Raman transitions between the qubit states.

For each atom, we set up the Hamiltonian in a rotating frame where the transformation frequencies match the respective laser frequencies driving the transitions of the atom we are describing. Within such a frame, the blue laser frequency component of the other atom appears as a term oscillating at a certain frequency. The individual Hamiltonian of atom $i$ ($i=1$ or $2$) is then:
\begin{widetext}
\begin{align}
H_i = & -(\Delta_{H,i} + \Delta_{b,i})|0\rangle\langle 0| 
-\Delta_{b,i}|1\rangle\langle 1| 
+ \Delta |e\rangle\langle e| 
- \Delta_r |r\rangle\langle r|
+ \frac{\Omega_{b,i}}{2} \Big( 
    e^{-i\phi_i} |0\rangle\langle e| 
  + e^{i\phi_i} |e\rangle\langle 0| 
  + e^{-i\phi_i} |1\rangle\langle e| 
  + e^{i\phi_i} |e\rangle\langle 1| 
\Big) \nonumber \\
&+ \frac{\Omega_{b,(3-i)}}{2} \Big( 
e^{-i\phi_{3-i} - i(\omega_{D, i}-\Delta_{b,i}+\Delta_{b,(3-i)}) t} |0\rangle\langle e| 
+ e^{i\phi_{3-i} + i(\omega_{D, i}-\Delta_{b,i}+\Delta_{b,(3-i)}) t} |e\rangle\langle 0|  \nonumber \\
&+ e^{-i\phi_{3-i} - i(\omega_{D, i}-\Delta_{b,i}+\Delta_{b,(3-i)}) t} |1\rangle\langle e| 
+ e^{i\phi_{3-i} + i(\omega_{D, i}-\Delta_{b,i}+\Delta_{b,(3-i)}) t} |e\rangle\langle 1| 
\Big) + \frac{\Omega_{r}}{2} \Big( |e\rangle\langle r| + |r\rangle\langle e| 
\Big)
\label{eq:Ham}
\end{align}
\end{widetext}
where we assume equal blue Rabi frequencies addressing the qubit states to the intermediate state. Here $\Delta_{H,i}$ is the qubit energy level splitting of atom $i$ (hyperfine splitting). $\Delta$ is the overall detuning from the intermediate state.
$\Delta_{b,i}$, $\Omega_{b,i}$ and $\phi_i$ are the additional detuning, Rabi frequency, and the phase of the blue laser frequency component corresponding to atom $i$, while $\Delta_{b,(3-i)}$, $\Omega_{b,(3-i)}$ and $\phi_{3-i}$ are the additional detuning, Rabi frequency, and the phase of the other blue laser frequency component. $\omega_{D, i}$ is the frequency of the $\ket{1}-\ket{e}$ transition of atom $i$ minus that of the other atom. $\Delta_r$ and $\Omega_r$ are the additional detuning and Rabi frequency of the red laser.

The Hamiltonians $H_i$ are $5\times 5$ in size, expressed in states $\ket{0}$, $\ket{1}$, $\ket{e}$, $\ket{r}$ and $\ket{t}$. $\ket{t}$ is not directly driven by the Hamiltonian, but is connected to $\ket{e}$ and $\ket{r}$ through jump operators, see Fig.~\ref{fig:levels}.  The Hamiltonian of the whole system can thus be written as:
\begin{equation}
    H = H_1\otimes I + I\otimes H_2 + V_{12}.
\label{equ:full_Ham}
\end{equation}

Here, $H_1$ and $H_2$ are the individual Hamiltonians for atoms 1 and 2, respectively. $V_{12}$ is the van der Waals interaction, expressed as
\begin{equation}
    V_{12} = B\ket{rr}\bra{rr}
\end{equation}
where $B$ is the interaction strength. The master equation of the system is
\begin{equation}
\dot{\rho} = -\frac{i}{\hbar}[H, \rho] 
+ \sum_k \gamma_k \left( 
    L_k \rho L_k^{\dagger} 
    - \frac{1}{2} \left\{ L_k^{\dagger} L_k, \rho \right\} 
\right).
\end{equation}
where the jump operators $L_{k}$ (and corresponding decay rates $\gamma_k$) account for the decay of Rydberg states, the dephasing of Rydberg states, and the decay of intermediate states.

In the limit of $\Delta \gg \Omega_b$, $\Delta \gg \Omega_r$, the two-photon process with two Rabi frequencies can be regarded as a one-photon process with an effective Rabi frequency $\Omega_{i}$ and an effective detuning $\delta_{\text{eff}}$ \cite{shore1990coherent}. Following a method analogous to that in \cite{PhysRevA.40.5103}, the effective Rabi frequency of such a system is:
\begin{equation}
    \Omega_{i} = \frac{\Omega_{b,i} \Omega_{r}}{2 (\Delta+\Delta_{b,i})}.
\end{equation}
For the purposes of our gate optimization as will follow, we choose to work with the same (dynamic) blue Rabi frequency for both atoms, while the red Rabi frequency remains unchanged. We set the maximal blue Rabi frequency to be $\Omega_{b, \text{max}}$. We choose to parameterize our gate with an overall effective two-photon Rabi frequency defined using the maximum blue Rabi frequency:
\begin{equation}
    \Omega_{\text{max}} := \frac{\Omega_{b,\text{max}} \Omega_{r}}{2 \Delta}.
    \label{equ:Rabi}
\end{equation}
The overall detuning of excitation to the Rydberg state for atom $i$ is:
\begin{align}
    \delta_{\text{eff},i} = \Delta_r-\Delta_{b,i}-\frac{\Omega_{b}^2}{4(\Delta+\Delta_{b,i})} 
         + \frac{\Omega_{r}^2}{4(\Delta+\Delta_{r})} \nonumber \\
         - \frac{\Omega_{b}^{\,2}}{4 (\Delta+\omega_{D,i}+\Delta_{b,(3-i)})}.
\label{equ:detuning}
\end{align}
The effective Rabi frequency here is the same as in a traditional two-photon driving model. However, different from that in a traditional two-photon driving model, the effective detuning has an additional term arising from the AC Stark shift induced by the second, off-resonant blue laser frequency component. This additional term distinguishes the dual-isotope model from a single-species model, where there is only one blue laser, and also from the dual-element model, where the wavelength difference between elements is large enough to render the influence of this term negligible. For the entangling gate, we adjust the laser detunings to set the effective detuning for each atom at the maximum blue Rabi frequency equal to zero.

\begin{table}[t]
\caption{\label{tab:para}%
Parameter Table for $^{85}$Rb and $^{87}$Rb}
\begin{ruledtabular}
\begin{tabular}{l dd}
\textrm{Parameter} &
\multicolumn{1}{c}{\textrm{$^{85}$Rb}} &
\multicolumn{1}{c}{\textrm{$^{87}$Rb}} \\
\hline
$\Delta/2\pi$/GHz        &  8        &  8        \\
$\Delta_H/2\pi$/GHz      &  3.036    &  6.835    \\
$\omega_{D}/2\pi$/GHz        &  1.142    & -1.142    \\
$\Omega_b/2\pi$/MHz      & 355.5     & 355.5       \\
$\Omega_r/2\pi$/MHz      & 454.5     & 454.5     \\
$V/2\pi$/MHz             & 200       & 200       \\
Rydberg $T_1$/$\mu$s     & 100       & 100       \\
Rydberg $T_2^*$/$\mu$s   &  6        &  6        \\
Intermediate $T_1$/ns    & 110       & 110       \\
\end{tabular}
\end{ruledtabular}
\end{table}

\begin{figure}[b]
\includegraphics[width=\linewidth]{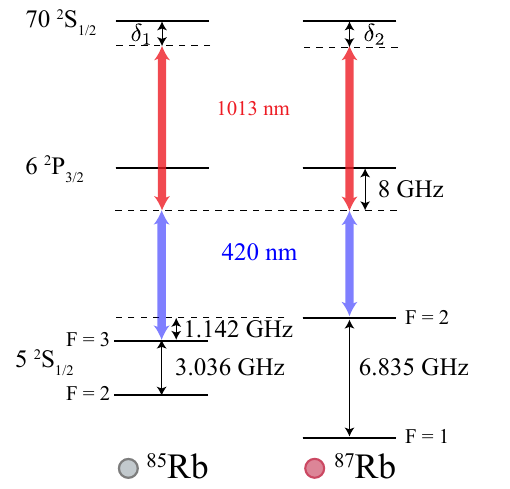}
\caption{Energy levels involved in the gate operation and key experimental parameters. Both atoms have $\ket{e} = \ket{6P_{3/2}}$ and $\ket{r} = \ket{70S_{1/2}}$, while $\ket{0}$ and $\ket{1}$ are different due to the difference in their hyperfine structure.}
\label{fig:levelRb}
\end{figure}

\begin{figure*}[htb]
\includegraphics[width=\linewidth]{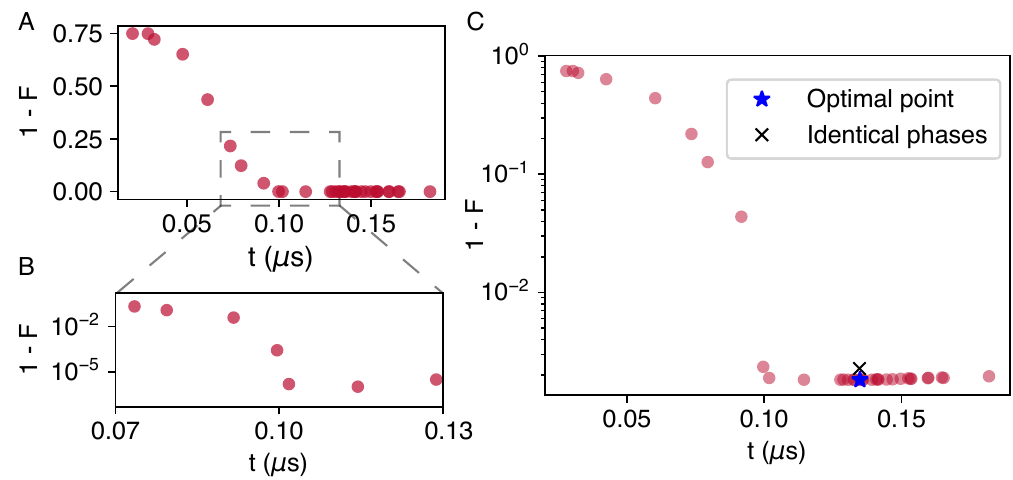}
\caption{\textbf{A. }Relation between unitary infidelity and time. Here, unitary infidelity refers to $1 - F$ from a Schrödinger equation simulation without non-unitary errors (such as decay and dephasing). Qubit state $\ket{0}$ is also excluded. The horizontal axis is the total gate time. \textbf{B. }Zoom in of the unitary infidelity between $t=0.07$~$\mu$s and $t=0.13$~$\mu$s. \textbf{C. }Relation between full infidelity and time. Here, full infidelity comes from $1 - F$ of a master equation simulation that includes dephasing and decay, representing a realistic experimental simulation. The optimal point incorporates independent phase functions for each atom (blue star). Optimizing the gate with an identical phase function for both atoms (up to a fixed phase offset $\varphi_{\text{diff}}$) results in a fidelity marked by $\times$.}
\label{fig:t-F_plot}
\end{figure*}


\subsection{High-Fidelity Two-Photon Driving Gate}

Many protocols have been proposed to implement two-qubit entangling gates with high fidelity and short interaction time \cite{PhysRevLett.85.2208, PhysRevLett.123.170503, PhysRevLett.100.170504, Jandura2022}. The first dual-isotope experiment utilized a $\pi$---2$\pi$---$\pi$ protocol \cite{zeng2017entangling}, as designed by \cite{PhysRevLett.85.2208}. Although this protocol is straightforward, high-fidelity operation is limited by the slow speed of the gate and undriven idling in the Rydberg state, increasing sensitivity to ground-Rydberg dephasing.

The high-fidelity entangling gate we build is based on the time-optimal gate from \cite{Jandura2022}. The time-optimal gate focuses on minimizing the duration of the entangling process with a fixed Rabi frequency by varying the phase of the laser in time. We consider this gate a good ansatz because shorter gate durations can mitigate errors arising from the decay and dephasing of the states. We incorporate the phase profile into our two-photon driving model through the phase of the blue lasers. Using an approximation analogous to \cite{Evered2023}, we model the phase profiles as simple sinusoids for convenience when implementing such gates into experimental platforms. Additionally, inspired by the smooth amplitude gate in \cite{Evered2023} to reduce intermediate-state scattering, we add an exponential rise and drop to the amplitude of the gate pulse. We find that only a small adjustment of parameters and a constant phase difference are required to build a high-fidelity two-photon pumping gate in our model.

The phase profiles are given by
\begin{equation}
    \phi_1(t) = A_1 \cos(\omega_1 t - \varphi_{1}) + \delta_{1} t + \varphi_{\text{diff}}
\end{equation}
\begin{equation}
    \phi_2(t) = A_2 \cos(\omega_2 t - \varphi_{2}) + \delta_{2} t
\end{equation}
and the pulse strength profile shared by both atoms is given by
\begin{equation}
    \Omega_b(t)=\Omega_{b,\text{max}} (1-e^{-t/T_r})(1-e^{-(T-t)/T_r}).
\end{equation}

Here, $\phi_1(t)$ is the phase modulation for atom 1 and $\phi_2(t)$ is the modulation for atom 2. $T$ is the total pulse length, $T_r$ is the rise and drop time, $\Omega_b$ is the shared blue Rabi frequency. Due to the dual-isotope nature of our system, we have three laser frequencies that can be adjusted (two blue frequency components and one red frequency component). There are two constraints to satisfy from setting the effective detuning to 0 in Eq.~\ref{equ:detuning} at the maximum Rabi frequency, leaving one extra degree of freedom. This degree of freedom is described by the detuning on the blue laser of atom 1, $\Delta_{b,1}$. Combining them together provides a high-fidelity gate described by 12 parameters: $A_1$, $A_2$, $\omega_1$, $\omega_2$, $\varphi_{1}$, $\varphi_{2}$, $\delta_{1}$, $\delta_{2}$, $\varphi_{\text{diff}}$, $T$, $T_r$ and $\Delta_{b,1}$.

\begin{figure*}[htb]
\includegraphics[width=\linewidth]{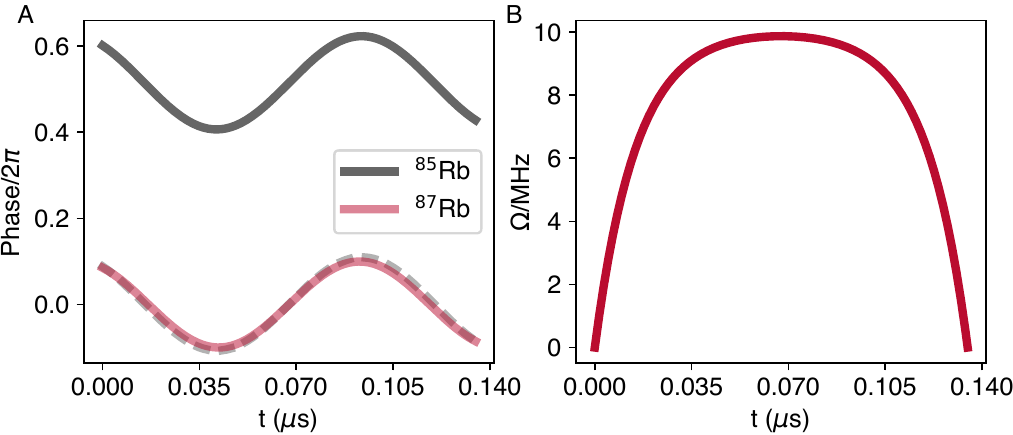}
\caption{\textbf{A. }Phase modulations for the $^{85}$Rb/$^{87}$Rb high-fidelity gate. The phase functions are not restricted in $[0, 2\pi)$ or $(-\pi, \pi]$ for a more intuitive graphical representation. To illustrate the difference between the two phase functions, we subtract the fixed phase offset from the phase function of $^{85}$Rb and plot it as the dashed grey line. \textbf{B. }Shared pulse strength profile for $^{85}$Rb/$^{87}$Rb high-fidelity gate. The vertical axis is calculated based on Eq.~\ref{equ:Rabi} but with varying blue Rabi frequency.}
\label{fig:phase}
\end{figure*}

\section{Simulation Based on $^{85}$Rb and $^{87}$Rb}

\subsection{Parameters for $^{85}$Rb and $^{87}$Rb}
For this section and the following, we set atom 1 as $^{85}$Rb and atom 2 as $^{87}$Rb. For $^{85}$Rb, $\ket{0}$ is $\ket{F=2, m_F = 0}$, $\ket{1}$ is $\ket{F=3, m_F = 0}$. For $^{87}$Rb, $\ket{0}$ is $\ket{F=1, m_F = 0}$, $\ket{1}$ is $\ket{F=2, m_F = 0}$. Both atoms have $\ket{e} = \ket{6P_{3/2}}$ and $\ket{r} = \ket{70S_{1/2}}$, see Fig.~\ref{fig:levelRb}.

For the simulation below, we assume both isotopes are excited to an $S$-orbital Rydberg state where the inter-atomic distance corresponds to a van der Waals interaction with an effective blockade strength of 200~MHz. Although it is also possible to build quantum gates with resonant dipolar interactions, a van der Waals interaction is used here because it provides a strong blockade at micron-scale distances, decays quicker as a function of inter-atomic distance, and possesses simpler energy level structures.

The parameters for each atom are listed in Table \ref{tab:para} and also illustrated in Fig.~\ref{fig:levelRb}. The Rabi frequencies are chosen to be 1.5 times the values experimentally demonstrated in \cite{Evered2023}. $T_1$ and $T_2^*$ times are estimated from \cite{Anand2024,Wurtz2023-yg}.  The decay rates $\gamma_k$ are derived from $T_1$ or $T_2^*$ assuming exponential decay or Gaussian decay, respectively.

\begin{figure}[h]
\includegraphics[width=\linewidth]{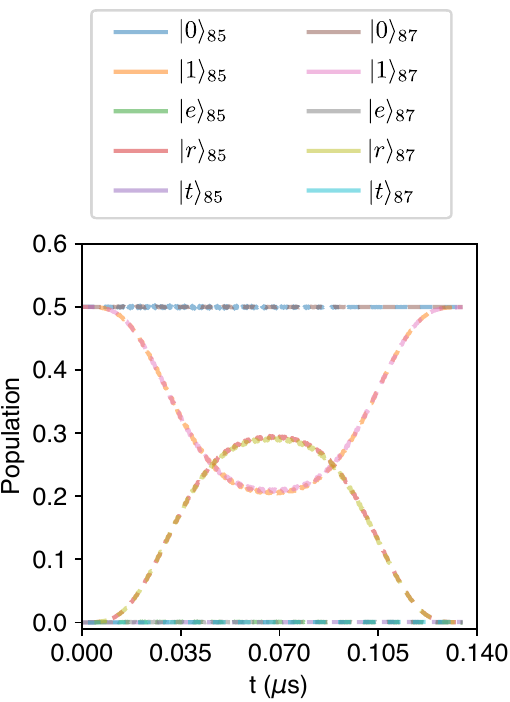}
\caption{Relation between state populations and time during the optimal gate.}
\label{fig:density}
\end{figure}

\subsection{Simulation Result}
We evaluate the fidelity of the entangling (CZ) gate by propagating the state $\frac{1}{2}(\ket{00}+\ket{01}+\ket{10}+\ket{11})$ with the system Hamiltonian and evaluating the Bell state fidelity $F$ at the end of the gate \cite{Jandura2022}. Numerical simulations are performed using the Quantum Toolbox in Python (QuTiP) \cite{qutip5}. To find the global maximum of the Bell-state fidelity, we use the Nelder-Mead method to search within the 12 parameter space for the lowest infidelity ($1-F$) in a fixed time window. We use a Schrödinger equation simulation that excludes the jump operators and the qubit state $\ket{0}$ of each atom when searching for the optimal gate at each time step. This reduces the simulation to only unitary dynamics and eases the computational demand required for the search. The result is shown in Fig.~\ref{fig:t-F_plot}A. As expected from the time-optimal ansatz, we observe a sharp drop in the infidelity beyond a minimal pulse duration (Fig.~\ref{fig:t-F_plot}B). We then evaluate the performance of the gate in a full master equation simulation that includes all decoherence channels. The result is shown in Fig.~\ref{fig:t-F_plot}C. The maximum fidelity is 0.9982 at $T=0.135$~$\mu$s, balancing the fidelity gain from a longer unitary gate against the increasing infidelity arising from decoherence over longer times, and surpassing the surface-code threshold \cite{PhysRevA.86.032324}.

For comparison, a traditional $\pi$---2$\pi$---$\pi$ gate with the same Rabi frequencies will only reach a Bell state fidelity of 0.9771 with a time duration of $T=0.198$~$\mu$s, notably longer than that of the optimal gate found in this work.

The parameters for the optimal gate are shown below. Here $A_1, \omega_1, \varphi_{1}, \delta_{1}$ are the parameters for $\textrm{$^{85}$Rb}$ and $A_2, \omega_2, \varphi_{2}, \delta_{2}$ are the parameters for $\textrm{$^{87}$Rb}$:
\[
\begin{aligned}
    A_1 &= 2\pi \times 0.1091, \quad & A_2 &= 2\pi \times 0.0998,\\
    \omega_1 &= 0.9407\, \Omega_{\text{max}}, \quad &  \omega_2 &= 0.9733  \, \Omega_{\text{max}},\\
    \varphi_{1} &= -0.6893, \quad & \varphi_{2} &=  -0.5489,\\
    \delta_{1} &= -0.0032 \, \Omega_{\text{max}}, \quad & \delta_{2} &=  0,\\
    T &=  1.0579 \times 2\pi/\Omega_{\text{max}} + 2 T_r, \quad & T_r &=  15.0316 \text{ ns},\\
    \Delta_{b,1} &= -0.6762 \text{ GHz},  \quad & \varphi_{\text{diff}} &=  3.2473,\\
\end{aligned}
\]
where $\Omega_{\text{max}}$ is the Rabi frequency calculated based on Eq.~\ref{equ:Rabi}. 

Fig.~\ref {fig:phase} plots the phase functions for both atoms and the pulse-shape amplitudes. In addition to the constant phase difference $\varphi_{\text{diff}}$, the phase functions of the two atoms differ slightly in shape. If we optimize for a gate that uses the same phase function for both atoms with a fixed phase offset $\varphi_{\text{diff}}$, we achieve a gate with a fidelity of $0.9977$ at the same time as the optimal gate with independent phase functions.
This shows that the gate infidelity is strongly reduced by using independent phase functions for each isotope.

The population of each state as a function of time during the gate is shown in Fig.~\ref{fig:density}. Population that is measured in the trash state can in principle be used to perform postselection or erasure conversion to further improve the fidelity of the gate \cite{Wu2022-ns}. 

Scattering from the intermediate states is the largest contribution to the infidelity of the optimal gate, accounting for 61\% of the error. Rydberg decay and dephasing account for 22\% and 13\% of the infidelity, respectively. The remaining 4\% is attributed to a coherent Raman transition in the qubit subspace arising from the presence of two frequency components in the blue laser, which can in principle be eliminated by including this process during gate optimization. 

These error sources immediately suggest strategies to improve the gate. We can increase the intermediate-state detuning to suppress intermediate-state scattering and increase the Rabi frequency to reduce the time spent in the Rydberg state \cite{Evered2026-cr}. For instance, by increasing the Rabi frequencies of the lasers and the intermediate-state detuning by a factor of two, we numerically find an optimal gate with a fidelity of 99.9\% and a gate time of 86.6 nanoseconds in a full master equation simulation. The laser intensities required to achieve these Rabi frequencies are now achievable in experimental platforms.
\\

\section{Conclusion and Outlook}
In conclusion, we report the design and simulation of a dual-isotope two-qubit entangling gate based on quantum optimal control. Such a two-qubit gate is one of the fundamental components of a universal quantum computer based on dual-isotope atom arrays. The simulation results show that the gate can maintain high fidelity and robust performance under the influence of blue-laser crosstalk, ensuring its unique advantage over traditional $\pi$---2$\pi$---$\pi$ gates even for near-detuned dual-isotope systems. Importantly, the gate error is below the surface-code threshold, allowing QEC based on existing error-correcting codes in such systems. The fidelity can be further improved using leakage detection, increased intermediate-state detuning, and higher effective Rabi frequency. The simulation here is based on the CZ gate in an $^{85}$Rb/$^{87}$Rb array, but the theoretical model applies for other dual-isotope arrays with a similar structure and also applies for the other entangling gates with reasonable adjustments. In addition, we demonstrated a general method to account for the influence of crosstalk when simulating dual-isotope systems. Considering the potential of dual-isotope atom arrays for qubit addressability, mid-circuit readout, and ancilla-based quantum error correction, we expect our work to be helpful for the current and next generation experimental platforms.

\begin{acknowledgments}
We thank Julian Sewell and Conor E. Bradley for fruitful discussions and critical reading of the manuscript. We acknowledge funding from the Air Force Office of Scientific Research (FA9550-26-1-B316). 
\end{acknowledgments}




\bibliography{bibliography}

@article{Jandura2022,
author = {Jandura, Sven and Pupillo, Guido},
year = {2022},
month = {05},
pages = {712},
title = {{Time-Optimal Two- and Three-Qubit Gates for Rydberg Atoms}},
volume = {6},
journal = {Quantum},
doi = {10.22331/q-2022-05-13-712}
}

@article{Evered2023,
  author    = {Simon J. Evered and Dolev Bluvstein and Marcin Kalinowski and Sepehr Ebadi and Tom Manovitz and Hengyun Zhou and Sophie H. Li and Alexandra A. Geim and Tout T. Wang and Nishad Maskara and Harry Levine and Giulia Semeghini and Markus Greiner and Vladan Vuleti{\'{c}} and Mikhail D. Lukin},
  title     = {{High-fidelity parallel entangling gates on a neutral-atom quantum computer}},
  journal   = {Nature},
  volume    = {622},
  number    = {7982},
  pages     = {268--272},
  year      = {2023},
  month     = {October},
  doi       = {10.1038/s41586-023-06481-y},
  url       = {https://doi.org/10.1038/s41586-023-06481-y},
  issn      = {1476-4687}
}

@inbook{shore1990coherent,
  author    = {B. W. Shore},
  title     = {{The Theory of Coherent Atomic Excitation}},
  chapter   = {13},
  volume    = {2},
  year      = {1990},
  publisher = {Wiley},
  address   = {New York}
}

@article{Ebadi2021,
  author    = {Sepehr Ebadi and Tout T. Wang and Harry Levine and Alexander Keesling and Giulia Semeghini and Ahmed Omran and Dolev Bluvstein and Rhine Samajdar and Hannes Pichler and Wen Wei Ho and Soonwon Choi and Subir Sachdev and Markus Greiner and Vladan Vuletić and Mikhail D. Lukin},
  title     = {{Quantum phases of matter on a 256-atom programmable quantum simulator}},
  journal   = {Nature},
  volume    = {595},
  number    = {7866},
  pages     = {227--232},
  year      = {2021},
  doi       = {10.1038/s41586-021-03582-4},
  url       = {https://doi.org/10.1038/s41586-021-03582-4},
  issn      = {1476-4687}
}

@article{Scholl2021,
  author    = {Pascal Scholl and Michael Schuler and Hannah J. Williams and Alexander A. Eberharter and Daniel Barredo and Kai-Niklas Schymik and Vincent Lienhard and Louis-Paul Henry and Thomas C. Lang and Thierry Lahaye and Andreas M. Läuchli and Antoine Browaeys},
  title     = {{Quantum simulation of 2D antiferromagnets with hundreds of Rydberg atoms}},
  journal   = {Nature},
  volume    = {595},
  number    = {7866},
  pages     = {233--238},
  year      = {2021},
  doi       = {10.1038/s41586-021-03585-1},
  url       = {https://doi.org/10.1038/s41586-021-03585-1},
  issn      = {1476-4687}
}

@article{Bluvstein2022,
  author    = {Dolev Bluvstein and Harry Levine and Giulia Semeghini and Tout T. Wang and Sepehr Ebadi and Marcin Kalinowski and Alexander Keesling and Nishad Maskara and Hannes Pichler and Markus Greiner and Vladan Vuletić and Mikhail D. Lukin},
  title     = {{A quantum processor based on coherent transport of entangled atom arrays}},
  journal   = {Nature},
  volume    = {604},
  number    = {7906},
  pages     = {451--456},
  year      = {2022},
  doi       = {10.1038/s41586-022-04592-6},
  url       = {https://doi.org/10.1038/s41586-022-04592-6},
  issn      = {1476-4687}
}

@article{Pause:24,
author = {Lars Pause and Lukas Sturm and Marcel Mittenb\"{u}hler and Stephan Amann and Tilman Preuschoff and Dominik Sch\"{a}ffner and Malte Schlosser and Gerhard Birkl},
journal = {Optica},
number = {2},
pages = {222--226},
publisher = {Optica Publishing Group},
title = {{Supercharged two-dimensional tweezer array with more than 1000 atomic qubits}},
volume = {11},
month = {Feb},
year = {2024},
url = {https://opg.optica.org/optica/abstract.cfm?URI=optica-11-2-222},
doi = {10.1364/OPTICA.513551},
}

@article{Werninghaus2021,
  author    = {Werninghaus, M. and Egger, D. J. and Roy, F. and Machnes, S. and Wilhelm, F. K. and Filipp, S.},
  title     = {{Leakage reduction in fast superconducting qubit gates via optimal control}},
  journal   = {npj Quantum Information},
  year      = {2021},
  volume    = {7},
  number    = {1},
  pages     = {14},
  doi       = {10.1038/s41534-020-00346-2},
  url       = {https://doi.org/10.1038/s41534-020-00346-2},
  issn      = {2056-6387},
}

@article{rembold2020introduction,
  title={{Introduction to quantum optimal control for quantum sensing with nitrogen-vacancy centers in diamond}},
  author={Rembold, Phila and Oshnik, Nimba and M{\"u}ller, Matthias M and Montangero, Simone and Calarco, Tommaso and Neu, Elke},
  journal={AVS Quantum Science},
  volume={2},
  number={2},
  year={2020},
  publisher={AIP Publishing}
}

@article{PhysRevLett.118.150503,
  title = {{Hybrid Quantum-Classical Approach to Quantum Optimal Control}},
  author = {Li, Jun and Yang, Xiaodong and Peng, Xinhua and Sun, Chang-Pu},
  journal = {Phys. Rev. Lett.},
  volume = {118},
  issue = {15},
  pages = {150503},
  numpages = {5},
  year = {2017},
  month = {Apr},
  publisher = {American Physical Society},
  doi = {10.1103/PhysRevLett.118.150503},
  url = {https://link.aps.org/doi/10.1103/PhysRevLett.118.150503}
}

@article{lis2023midcircuit,
  title={{Midcircuit operations using the omg architecture in neutral atom arrays}},
  author={Lis, Joanna W and Senoo, Aruku and McGrew, William F and R{\"o}nchen, Felix and Jenkins, Alec and Kaufman, Adam M},
  journal={Physical Review X},
  volume={13},
  number={4},
  pages={041035},
  year={2023},
  publisher={APS}
}

@article{singh2023mid,
  title={{Mid-circuit correction of correlated phase errors using an array of spectator qubits}},
  author={Singh, Kevin and Bradley, Conor E and Anand, Shraddha and Ramesh, Vikram and White, Ryan and Bernien, Hannes},
  journal={Science},
  volume={380},
  number={6651},
  pages={1265--1269},
  year={2023},
  publisher={American Association for the Advancement of Science}
}

@article{bluvstein2024logical,
  title={{Logical quantum processor based on reconfigurable atom arrays}},
  author={Bluvstein, Dolev and Evered, Simon J and Geim, Alexandra A and Li, Sophie H and Zhou, Hengyun and Manovitz, Tom and Ebadi, Sepehr and Cain, Madelyn and Kalinowski, Marcin and Hangleiter, Dominik and others},
  journal={Nature},
  volume={626},
  number={7997},
  pages={58--65},
  year={2024},
  publisher={Nature Publishing Group UK London}
}

@article{norcia2023midcircuit,
  title={{Midcircuit qubit measurement and rearrangement in a Yb 171 atomic array}},
  author={Norcia, MA and Cairncross, WB and Barnes, K and Battaglino, P and Brown, A and Brown, MO and Cassella, K and Chen, C-A and Coxe, R and Crow, D and others},
  journal={Physical Review X},
  volume={13},
  number={4},
  pages={041034},
  year={2023},
  publisher={APS}
}

@article{beterov2015rydberg,
  title={{Rydberg blockade, F{\"o}rster resonances, and quantum state measurements with different atomic species}},
  author={Beterov, II and Saffman, M},
  journal={Physical Review A},
  volume={92},
  number={4},
  pages={042710},
  year={2015},
  publisher={APS}
}

@article{zeng2017entangling,
  title={{Entangling two individual atoms of different isotopes via Rydberg blockade}},
  author={Zeng, Yong and Xu, Peng and He, Xiaodong and Liu, Yangyang and Liu, Min and Wang, Jin and Papoular, DJ and Shlyapnikov, GV and Zhan, Mingsheng},
  journal={Physical Review Letters},
  volume={119},
  number={16},
  pages={160502},
  year={2017},
  publisher={APS}
}

@article{sheng2022defect,
  title={{Defect-free arbitrary-geometry assembly of mixed-species atom arrays}},
  author={Sheng, Cheng and Hou, Jiayi and He, Xiaodong and Wang, Kunpeng and Guo, Ruijun and Zhuang, Jun and Mamat, Bahtiyar and Xu, Peng and Liu, Min and Wang, Jin and others},
  journal={Physical Review Letters},
  volume={128},
  number={8},
  pages={083202},
  year={2022},
  publisher={APS}
}

@article{PhysRevA.40.5103,
  title = {{Experimental study of one- and two-photon Rabi oscillations}},
  author = {Gentile, Thomas R. and Hughey, Barbara J. and Kleppner, Daniel and Ducas, Theodore W.},
  journal = {Phys. Rev. A},
  volume = {40},
  issue = {9},
  pages = {5103--5115},
  numpages = {0},
  year = {1989},
  month = {Nov},
  publisher = {American Physical Society},
  doi = {10.1103/PhysRevA.40.5103},
  url = {https://link.aps.org/doi/10.1103/PhysRevA.40.5103}
}

@article{PhysRevA.86.032324,
  title = {{Surface codes: Towards practical large-scale quantum computation}},
  author = {Fowler, Austin G. and Mariantoni, Matteo and Martinis, John M. and Cleland, Andrew N.},
  journal = {Phys. Rev. A},
  volume = {86},
  issue = {3},
  pages = {032324},
  numpages = {48},
  year = {2012},
  month = {Sep},
  publisher = {American Physical Society},
  doi = {10.1103/PhysRevA.86.032324},
  url = {https://link.aps.org/doi/10.1103/PhysRevA.86.032324}
}

@article{PhysRevLett.85.2208,
  title = {{Fast Quantum Gates for Neutral Atoms}},
  author = {Jaksch, D. and Cirac, J. I. and Zoller, P. and Rolston, S. L. and C\^ot\'e, R. and Lukin, M. D.},
  journal = {Phys. Rev. Lett.},
  volume = {85},
  issue = {10},
  pages = {2208--2211},
  numpages = {0},
  year = {2000},
  month = {Sep},
  publisher = {American Physical Society},
  doi = {10.1103/PhysRevLett.85.2208},
  url = {https://link.aps.org/doi/10.1103/PhysRevLett.85.2208}
}

@article{PhysRevLett.123.170503,
  title = {{Parallel Implementation of High-Fidelity Multiqubit Gates with Neutral Atoms}},
  author = {Levine, Harry and Keesling, Alexander and Semeghini, Giulia and Omran, Ahmed and Wang, Tout T. and Ebadi, Sepehr and Bernien, Hannes and Greiner, Markus and Vuleti\ifmmode \acute{c}\else \'{c}\fi{}, Vladan and Pichler, Hannes and Lukin, Mikhail D.},
  journal = {Phys. Rev. Lett.},
  volume = {123},
  issue = {17},
  pages = {170503},
  numpages = {6},
  year = {2019},
  month = {Oct},
  publisher = {American Physical Society},
  doi = {10.1103/PhysRevLett.123.170503},
  url = {https://link.aps.org/doi/10.1103/PhysRevLett.123.170503}
}

@article{PhysRevLett.100.170504,
  title = {{Quantum Gates and Multiparticle Entanglement by Rydberg Excitation Blockade and Adiabatic Passage}},
  author = {M\o{}ller, Ditte and Madsen, Lars Bojer and M\o{}lmer, Klaus},
  journal = {Phys. Rev. Lett.},
  volume = {100},
  issue = {17},
  pages = {170504},
  numpages = {4},
  year = {2008},
  month = {Apr},
  publisher = {American Physical Society},
  doi = {10.1103/PhysRevLett.100.170504},
  url = {https://link.aps.org/doi/10.1103/PhysRevLett.100.170504}
}

@article{PhysRevX.12.011040,
  title = {{Dual-Element, Two-Dimensional Atom Array with Continuous-Mode Operation}},
  author = {Singh, Kevin and Anand, Shraddha and Pocklington, Andrew and Kemp, Jordan T. and Bernien, Hannes},
  journal = {Phys. Rev. X},
  volume = {12},
  issue = {1},
  pages = {011040},
  numpages = {11},
  year = {2022},
  month = {Mar},
  publisher = {American Physical Society},
  doi = {10.1103/PhysRevX.12.011040},
  url = {https://link.aps.org/doi/10.1103/PhysRevX.12.011040}
}

@article{PhysRevA.105.063111,
  title = {{Simple, passive design for large optical trap arrays for single atoms}},
  author = {Huft, P. and Song, Y. and Graham, T. M. and Jooya, K. and Deshpande, S. and Fang, C. and Kats, M. and Saffman, M.},
  journal = {Phys. Rev. A},
  volume = {105},
  issue = {6},
  pages = {063111},
  numpages = {12},
  year = {2022},
  month = {Jun},
  publisher = {American Physical Society},
  doi = {10.1103/PhysRevA.105.063111},
  url = {https://link.aps.org/doi/10.1103/PhysRevA.105.063111}
}

@article{2ym8-vs82,
  title = {{AI-Enabled Parallel Assembly of Thousands of Defect-Free Neutral Atom Arrays}},
  author = {Lin, Rui and Zhong, Han-Sen and Li, You and Zhao, Zhang-Rui and Zheng, Le-Tian and Hu, Tai-Ran and Wu, Hong-Ming and Wu, Zhan and Ma, Wei-Jie and Gao, Yan and Zhu, Yi-Kang and Su, Zhao-Feng and Ouyang, Wan-Li and Zhang, Yu-Chen and Rui, Jun and Chen, Ming-Cheng and Lu, Chao-Yang and Pan, Jian-Wei},
  journal = {Phys. Rev. Lett.},
  volume = {135},
  issue = {6},
  pages = {060602},
  numpages = {7},
  year = {2025},
  month = {Aug},
  publisher = {American Physical Society},
  doi = {10.1103/2ym8-vs82},
  url = {https://link.aps.org/doi/10.1103/2ym8-vs82}
}

@article{Anand2024,
  author       = {Shraddha Anand and Conor E. Bradley and Ryan White and Vikram Ramesh and Kevin Singh and Hannes Bernien},
  title        = {{A dual-species Rydberg array}},
  journal = {Nature Physics},
  year         = {2024},
  volume       = {20},
  number       = {11},
  pages        = {1744--1750},
  doi          = {10.1038/s41567-024-02638-2},
  url          = {https://doi.org/10.1038/s41567-024-02638-2},
  issn         = {1745-2481},
  date         = {2024-11-01},
}

@article{chen2023continuous,
  title={{Continuous symmetry breaking in a two-dimensional Rydberg array}},
  author={Chen, Cheng and Bornet, Guillaume and Bintz, Marcus and Emperauger, Gabriel and Leclerc, Lucas and Liu, Vincent S and Scholl, Pascal and Barredo, Daniel and Hauschild, Johannes and Chatterjee, Shubhayu and others},
  journal={Nature},
  volume={616},
  number={7958},
  pages={691--695},
  year={2023},
  publisher={Nature Publishing Group UK London}
}

@article{shaw2024multi,
  title={{Multi-ensemble metrology by programming local rotations with atom movements}},
  author={Shaw, Adam L and Finkelstein, Ran and Tsai, Richard Bing-Shiun and Scholl, Pascal and Yoon, Tai Hyun and Choi, Joonhee and Endres, Manuel},
  journal={Nature Physics},
  volume={20},
  number={2},
  pages={195--201},
  year={2024},
  publisher={Nature Publishing Group UK London}
}

@misc{miles2026qubitsyndromemeasurementshigh,
      title={{Qubit syndrome measurements with a high fidelity Rb-Cs Rydberg gate}}, 
      author={J. Miles and M. T. Lichtman and A. M. Scott and J. Scott and S. A. Norrell and M. J. Bedalov and D. A. Belknap and D. C. Cole and S. Y. Eubanks and M. Gillette and P. Gokhale and J. Goldwin and M. Iliev and R. A. Jones and K. W. Kuper and D. Mason and P. T. Mitchell and J. D. Murphree and N. A. Neff-Mallon and T. W. Noel and A. G. Radnaev and I. V. Vinogradov and M. Saffman},
      year={2026},
      eprint={2603.13492},
      archivePrefix={arXiv},
      primaryClass={quant-ph},
      url={https://arxiv.org/abs/2603.13492}, 
}

@ARTICLE{Wu2022-ns,
  title     = "{Erasure conversion for fault-tolerant quantum computing in
               alkaline earth Rydberg atom arrays}",
  author    = "Wu, Yue and Kolkowitz, Shimon and Puri, Shruti and Thompson,
               Jeff D",
  journal   = "Nat. Commun.",
  publisher = "Springer Science and Business Media LLC",
  volume    =  13,
  number    =  1,
  pages     = "4657",
  month     =  aug,
  year      =  2022,
  copyright = "https://creativecommons.org/licenses/by/4.0"
}

@article{PhysRevX.14.041062,
  title = {{Hybrid Atom Tweezer Array of Nuclear Spin and Optical Clock Qubits}},
  author = {Nakamura, Yuma and Kusano, Toshi and Yokoyama, Rei and Saito, Keito and Higashi, Koichiro and Ozawa, Naoya and Takano, Tetsushi and Takasu, Yosuke and Takahashi, Yoshiro},
  journal = {Phys. Rev. X},
  volume = {14},
  issue = {4},
  pages = {041062},
  numpages = {14},
  year = {2024},
  month = {Dec},
  publisher = {American Physical Society},
  doi = {10.1103/PhysRevX.14.041062},
  url = {https://link.aps.org/doi/10.1103/PhysRevX.14.041062}
}

@misc{Wurtz2023-yg,
  title         = "{Aquila: {QuEra's} 256-qubit neutral-atom quantum computer}",
  author        = "Wurtz, Jonathan and Bylinskii, Alexei and Braverman, Boris
                   and Amato-Grill, Jesse and Cantu, Sergio H and Huber,
                   Florian and Lukin, Alexander and Liu, Fangli and Weinberg,
                   Phillip and Long, John and Wang, Sheng-Tao and Gemelke,
                   Nathan and Keesling, Alexander",
  month         =  jun,
  year          =  2023,
  copyright     = "http://arxiv.org/licenses/nonexclusive-distrib/1.0/",
  archivePrefix = "arXiv",
  primaryClass  = "quant-ph",
  eprint        = "2306.11727"
}

@ARTICLE{Li2026-ql,
  title     = "{Fast, continuous, and coherent atom reloading in a neutral-atom
               qubit array}",
  author    = "Li, Yiyi and Bao, Yicheng and Peper, Michael and Li, Chenyuan
               and Thompson, Jeff D",
  journal   = "Science",
  publisher = "American Association for the Advancement of Science (AAAS)",
  volume    =  393,
  number    =  6817,
  pages     = "1213--1216",
  month     =  sep,
  year      =  2026,
  copyright = "https://www.science.org/content/page/science-licenses-journal-article-reuse"
}

@article{qutip5,
  title = {{QuTiP 5: The Quantum Toolbox in {Python}}},
  author = {
    Lambert, Neill and Gigu{`e}re, Eric and Menczel, Paul and Li, Boxi and
    Hopf, Patrick and Su{'a}rez, Gerardo and Gali, Marc and Lishman, Jake and
    Gadhvi, Rushiraj and Agarwal, Rochisha and Galicia, Asier and Shammah, Nathan and
    Nation, Paul and Johansson, J. R. and Ahmed, Shahnawaz and Cross, Simon and
    Pitchford, Alexander and Nori, Franco
  },
  journal = {Physics Reports},
  volume = {1153},
  pages = {1-62},
  year = {2026},
  issn = {0370-1573},
  doi = {10.1016/j.physrep.2025.10.001},
  url = {https://www.sciencedirect.com/science/article/pii/S0370157325002704},
}

@ARTICLE{Chiu2025-eg,
  title     = "{Continuous operation of a coherent 3,000-qubit system}",
  author    = "Chiu, Neng-Chun and Trapp, Elias C and Guo, Jinen and Abobeih,
               Mohamed H and Stewart, Luke M and Hollerith, Simon and
               Stroganov, Pavel L and Kalinowski, Marcin and Geim, Alexandra A
               and Evered, Simon J and Li, Sophie H and Lyu, Xingjian and
               Peters, Lisa M and Bluvstein, Dolev and Wang, Tout T and
               Greiner, Markus and Vuleti{\'c}, Vladan and Lukin, Mikhail D",
  journal   = "Nature",
  publisher = "Springer Science and Business Media LLC",
  volume    =  646,
  number    =  8087,
  pages     = "1075--1080",
  month     =  oct,
  year      =  2025,
  copyright = "https://creativecommons.org/licenses/by-nc-nd/4.0"
}

@misc{bhardwaj2026highrateqldpcprocessors,
      title={{High-rate qLDPC processors}}, 
      author={Aditya Bhardwaj and Muzhou Ma and Nadine Meister and Robbie King and Dolev Bluvstein and John Preskill and Madelyn Cain and Qian Xu and Hsin-Yuan Huang},
      year={2026},
      eprint={2607.28795},
      archivePrefix={arXiv},
      primaryClass={quant-ph},
      url={https://arxiv.org/abs/2607.28795}, 
}

@article{PhysRevLett.129.203602,
  title = {{Mid-Circuit Cavity Measurement in a Neutral Atom Array}},
  author = {Deist, Emma and Lu, Yue-Hui and Ho, Jacquelyn and Pasha, Mary Kate and Zeiher, Johannes and Yan, Zhenjie and Stamper-Kurn, Dan M.},
  journal = {Phys. Rev. Lett.},
  volume = {129},
  issue = {20},
  pages = {203602},
  numpages = {7},
  year = {2022},
  month = {Nov},
  publisher = {American Physical Society},
  doi = {10.1103/PhysRevLett.129.203602},
  url = {https://link.aps.org/doi/10.1103/PhysRevLett.129.203602}
}

@ARTICLE{graham2023midcircuit,
  title     = "{Midcircuit measurements on a single-species neutral alkali atom
               quantum processor}",
  author    = "Graham, T M and Phuttitarn, L and Chinnarasu, R and Song, Y and
               Poole, C and Jooya, K and Scott, J and Scott, A and Eichler, P
               and Saffman, M",
  journal   = "Phys. Rev. X.",
  publisher = "American Physical Society (APS)",
  volume    =  13,
  number    =  4,
  month     =  dec,
  year      =  2023,
  copyright = "https://creativecommons.org/licenses/by/4.0/"
}

@ARTICLE{Evered2026-cr,
  title         = "{High-fidelity entangling gates and nonlocal circuits with
                   neutral atoms}",
  author        = "Evered, Simon J and Xu, Muqing and Li, Sophie H and Geim,
                   Alexandra A and Pablo Bonilla Ataides, J and Kalinowski,
                   Marcin and Bluvstein, Dolev and Maskara, Nishad and Kokail,
                   Christian and Greiner, Markus and Vuleti{\'c}, Vladan and
                   Lukin, Mikhail D",
    journal = "",
  month         =  apr,
  year          =  2026,
  copyright     = "http://arxiv.org/licenses/nonexclusive-distrib/1.0/",
  archivePrefix = "arXiv",
  primaryClass  = "quant-ph",
  eprint        = "2604.25987"
}

\end{document}